\documentclass[manuscript,screen, nonacm]{acmart}

\AtBeginDocument{%
  \providecommand\BibTeX{{%
    \normalfont B\kern-0.5em{\scshape i\kern-0.25em b}\kern-0.8em\TeX}}}

\setcopyright{acmcopyright}
\copyrightyear{2018}
\acmYear{2018}
\acmDOI{XXXXXXX.XXXXXXX}

\acmConference[Conference acronym 'XX]{Make sure to enter the correct
  conference title from your rights confirmation emai}{June 03--05,
  2018}{Woodstock, NY}

\acmBooktitle{Woodstock '18: ACM Symposium on Neural Gaze Detection,
 June 03--05, 2018, Woodstock, NY} 
\acmPrice{15.00}
\acmISBN{978-1-4503-XXXX-X/18/06}

\usepackage{algorithm}
\usepackage{algorithmic}

\usepackage{xcolor}
\usepackage{comment}
\usepackage{booktabs}

\usepackage{newfloat}
\usepackage{listings}
\floatstyle{ruled}
\newfloat{listing}{tb}{lst}{}
\floatname{listing}{Listing}

\definecolor{green(ncs)}{rgb}{0.0, 0.62, 0.42}
\definecolor{green(pigment)}{rgb}{0.0, 0.65, 0.31}
\definecolor{forestgreen(web)}{rgb}{0.13, 0.55, 0.13}
\definecolor{Orange}{rgb}{1,0.5,0}
\definecolor{Red}{rgb}{1,0,0}
 \definecolor{Green}{rgb}{0,0.8,0.5}
\definecolor{Purple}{rgb}{0.75,0,1}
\definecolor{babypink}{rgb}{0.96, 0.76, 0.76}
\definecolor{azure}{rgb}{0,0.49,1}
\definecolor{periwinkle}{rgb}{0.8, 0.8, 1.0}
\definecolor{Pink}{RGB}{255, 102, 204}
\definecolor{electriccyan}{rgb}{0.0, 1.0, 1.0}
\definecolor{dodgerblue}{rgb}{0.12, 0.56, 1.0}

\newcommand{\angelica}[1]{\textsf{\textbf{\textcolor{Green}{[[AG: #1]]}}}}

\begin{document}
\title{Likes and Fragments: Examining Perceptions of Time Spent on TikTok}

\author{Angelica Goetzen}
\authornote{Both authors contributed equally to this research.}
\affiliation{%
  \institution{Max Planck Institute for Software Systems}
  \city{Saarbrucken}
  \country{Germany}
}

\author{Ruizhe Wang}
\authornote{This work was done at MPI-SWS.}
\authornotemark[1]
\affiliation{%
  \institution{University of Waterloo}
  \city{Waterloo}
  \country{Canada}
}

\author{Elissa M. Redmiles}
\affiliation{%
  \institution{Max Planck Institute for Software Systems}
  \city{Saarbrucken}
  \country{Germany}
}

\author{Savvas Zannettou}
\affiliation{%
  \institution{TU Delft}
  \city{Delft}
  \country{Netherlands}
}

\author{Oshrat Ayalon}
\affiliation{%
  \institution{Max Planck Institute for Software Systems}
  \city{Saarbrucken}
  \country{Germany}
}


    
    

\begin{abstract}
Researchers use information about the amount of time people spend on digital media for numerous purposes. While social media platforms commonly do not allow external access to measure the use time directly, a usual alternative method is to use participants' self-estimation. However, doubts were raised about the self-estimation's accuracy, posing questions regarding the cognitive factors that underline people's perceptions of the time they spend on social media. In this work, we build on prior studies and explore a novel social media platform in the context of use time: TikTok. We conduct platform-independent measurements of people's self-reported and server-logged TikTok usage (n=255) to understand how users' demographics and platform engagement influence their perceptions of the time they spend on the platform and their estimation accuracy. Our work adds to the body of work seeking to understand time estimations in different digital contexts and identifies new influential engagement factors.

\end{abstract}


\maketitle
\section{Introduction}

Assessing users' digital media use has been of interest to researchers across disciplines. Digital media use has been connected to users' psychological well-being (e.g., \citet{valkenburg2021social,schonning2020social}), physical well-being (e.g., \citet{orzech2016digital,zeeni2018media}), and cognitive processes like attention and learning (e.g., \citet{ra2018association}). The popularity of digital platforms like social media underpins the importance of a deeper understanding of the implications of digital media use. Measuring digital media use accurately, however, remains a challenge. 

Most digital media research relies on self-reported estimates of usage by participants \cite{griffioen2020toward}. Yet, participants' self-reports of their time spent on digital media are frequently inaccurate \cite{parry2021systematic}. Little is known about the cognitive processes behind this phenomenon, with psychology and media scholars speculating that factors like memory \cite{schwarz2001asking,larson2006predicting}, reporting bias from personal and societal views of social media \cite{lee2021role,junco2013comparing,Podsakoff2003}, or interpretability issues within question wording or other aspects of study design \cite{schwarz2001asking,Ernala_2020,junco2013comparing,mieczkowski2020priming} play a role. Newer technologies like screentime trackers, which allow researchers to compare participants' self-reports of their time spent using digital media to their actual logged usage, have shed light on the inaccuracies of these self-reports; prior work has found that users inaccurately report their time spent on the internet \cite{Scharkow2016,araujo2017much}, using mobile phones \cite{Sewall2020,ellis2019smartphone,Ohme2020}, and, notably, on social media platforms \cite{mahalingham2023assessing,coyne2023comparison,Sewall2020,verbeij2021accuracy,Ernala_2020,verbeij2022experience,Boyle2022,Burnell2021,junco2013comparing,rozgonjuk2020instagram}. This kind of logged data cannot be obtained by researchers externally, and must be shared by participants. Through analysis of self-reported time estimations and logged usage data, researchers have determined potentially relevant demographic, usage, and context-based factors that impact users' self-reporting accuracy.

Though time estimation research has been explored in the context of different social media platforms such as Facebook, Snapchat, and Instagram, research has yet to explore how users estimate their time spent on TikTok. TikTok, a highly popular social media platform \cite{demandsage:online}, differentiates itself from other platforms by centering the user experience almost entirely on short-form video content \cite{forbes3}. The content on TikTok is selected and served to users via a content recommendation system, which 
users find uniquely compelling \cite{Bhandari_2022,klug2021trick}.
Usage of short-form video platforms is a relatively novel variable of study, distinguishable from types of social media use studied in the past \cite{chao2023tiktok}; it is, therefore, important to extend prior work on social media time estimation to cover platforms whose interfaces may have different implications for users than platforms previously studied in a time-estimation context.

We add to the body of work on social media usage research by investigating how TikTok users estimate their time spent on the short-form video platform, and compare those estimations with actual logged usage to identify inaccuracies. We then examine relationships between the accuracy of self-reports and demographic and user behavior factors; going beyond the engagement measures in prior work, which were limited to the users' tenure on the platform and the number of sessions of engagement per day~\cite{Ernala_2020}, we explore a richer set of platform behavior and engagement metrics such as likes and number of videos watched. 

We leverage a previously collected dataset~\cite{blind} that contains TikTok user data packages donated by 255 TikTok users. This dataset includes information on users' interactions with the TikTok platform, like their logged usage time, number of watched videos, and their engagement with videos. It also contains self-reported survey data, from which we draw participants' demographic information and their self-reported estimates of their time spent on TikTok. Through our analysis, we find that 
overall, 
participants overestimate the time they spend on TikTok. Further, their estimates of time spent are not correlated with their actual logged usage time on TikTok. 
We identify predictors of inaccuracies among our participants; first, we find that those who engage more with the platform by liking more videos, estimate that they spend longer time on TikTok (and the error in their estimations is higher). 
Second, we find that the number of sessions, or times participants opened TikTok, is negatively related to their estimates of time spent (a higher amount of sessions means a lower estimate of time spent), and positively associated with accuracy (a higher amount of sessions means less over-estimation error). 

This work provides the first evaluation of the accuracy of self-reported time on a short-form video platform (TikTok) to find that participants generally overestimate their use on TikTok. We provide possible theories based on social media and cognitive psychology literature to explain the correlations between the self-reported time and engagement, and provide suggestions for future study design.

\section{Related Work}
\label{sec:related}
Here, we summarize prior work on the accuracy of self-reports of social media use, offer background on the TikTok platform and how we account for its unique features in our analysis, and briefly summarize prior work on data donation methodology. 

\subsection{Accuracy of Self-Reports of Social Media Use}

A recent meta-analysis of research on self-reports of time spent on digital media found only a moderate association between self-reported digital media usage and logged usage across studies \cite{parry2021systematic}. Indeed, work utilizing recent technological advances allowing researchers to view users' actual logged usage times suggests that asking participants to assess their time spent using \textit{social media,} the digital media of focus in our study, is an unreliable means of measurement. Prior work finds that participants have a tendency to overestimate the time they spend on social media \cite{coyne2023comparison,
Burnell2021,Ernala_2020,verbeij2021accuracy,Sewall2020,junco2013comparing}, yet little is known about what causes inaccuracies between user estimates and actual usage. Researchers have identified user demographics and engagement factors correlated with accuracy; for instance, individuals who spend more time on social media platforms estimate their time less accurately \cite{Ernala_2020,Boyle2022,Sewall2020}. Self-report accuracy may also be context-dependent, as differences in the accuracy of estimations have been noted across social media platforms \cite{verbeij2021accuracy,Ernala_2020,Burnell2021}. 

Further, prior work has posited potential cognitive explanations for inaccuracies in self-reporting. 
Participants may struggle to accurately recall their behaviors \cite{schwarz2001asking}; both features of the activity (like its complexity or novelty) and users' participation (like their emotional and intellectual engagement) can impact their perception of their time spent \cite{larson2006predicting}. 
For social media use in particular, use can be sporadic and fragmented throughout the day, or multi-tasked with other activities \cite{voorveld2013age,voorveld2014investigating}, which adds a unique challenge to estimating total usage time \cite{verbeij2021accuracy}. Participants may also report their behaviors based on perceptions of themselves \cite{schwarz2001asking}, attitudes towards social media \cite{lee2021role,junco2013comparing,Podsakoff2003}, or a desire to appear a certain way \cite{latkin2017relationship}, rather than based on reality. Additionally, inaccuracies can stem from participants' varied interpretations of the research design, i.e., some may interpret the wording and framing of self-report questions differently than others or than the study intends \cite{schwarz2001asking,Ernala_2020,junco2013comparing,mieczkowski2020priming}. 

Researchers have attempted to optimize research designs and methodologies to capture the time users spend on social media more accurately (e.g., \cite{jenkins2013development,mieczkowski2020priming,block2018prospective}). However, the widespread inaccuracies across self-report measures warrant a deeper exploration into users' habits of time estimation, and what factors correlate with inaccuracies. We extend the literature in this area by assessing how users report their time spent on TikTok, a social media platform whose users have not yet been singularly investigated on their self-report behaviors.

\subsection{TikTok}

Prior work has studied users' perceptions of their time spent on social media in aggregate, and in the context of specific platforms like Facebook and Instagram. Short-video application usage (SVU) is a newer variable of study \cite{chao2023tiktok}, potentially distinct from the broader variable of social media use; placing this variable at the center of our analysis, our work focuses on TikTok, a prominent social media platform whose interface famously centers the user experience on short-form video content.


When using TikTok, users may scroll through two different content feeds, one containing short-form videos posted by the people they follow (“Following”), the other an endlessly scrollable, curated feed of content from different creators (“For You”). 
While TikTok users can engage with classic social media features like messaging and commenting, the highly-personalized short-form video recommendation system on the For You feed is arguably its most distinguishable feature \cite{nyt1}, playing a large role in continued user engagement with the app \cite{Bhandari_2022,klug2021trick}.
In recent years, platforms like Instagram have integrated feeds of short-form content into their interfaces following the success of TikTok \cite{wired1}.

As self-reporting inaccuracies are seen to vary across social media platforms already \cite{verbeij2021accuracy,Ernala_2020}, we consider TikTok's distinct content landscape and uniquely engaged user base and investigate 
how users retrospectively estimate the time they spend on the app. 
 In addition to other factors, we measure forms of user engagement, such as the number of videos they watch and the number of videos they ``like,'' to understand how engagement may influence users's estimates of their time spent on this platform.

\subsection{Data Donation}

Per Article 15 of the  EU's General Data Protection Regulation ~\cite{gdpr}, which describes rights of access for data subjects, most major digital platforms now provide their users with electronic access to the personal data they collect and process for each user via downloadable data packages \cite{boeschoten2020digital}. Researchers studying digital media platforms and user behavior are beginning to leverage the rich information in these packages by requesting that users donate them for study \cite{van2021promises,wei2020twitter,baumgartner2022novel}. 
To assess the accuracy of users' estimations of their time spent on TikTok, and to explore other user factors that may correlate with logged usage time, we compare self-reported time estimations collected from a survey against logged usage data using a dataset~\cite{blind} containing donated data, as detailed in Methodology.

\section{Methodology}
\label{sec:collection}

\begin{table*}[t!]
    \centering
\begin{tabular}{@{}lrlrlr@{}}
\toprule
\multicolumn{2}{c}{\textbf{Gender}}     & \multicolumn{2}{c}{\textbf{Age}}   & \multicolumn{2}{c}{\textbf{Education}}        \\ \midrule
Woman & \multicolumn{1}{r|}{120 (47\%)} & 18-24 years old & \multicolumn{1}{r|}{118 (46\%)} & Below bachelor's degree  & 112 (44\%) \\
Man   & \multicolumn{1}{r|}{135 (53\%)} & 25-64 years old & \multicolumn{1}{r|}{137 (54\%)} & Holds a bachelor's degree or above & 143 (56\%) \\ \bottomrule
\end{tabular}
    \caption{Demographics of survey participants.}
    \label{tab:demographics}
\end{table*}

\begin{table}[t]
\centering
\begin{tabular}{@{}lrrrrr@{}}
\toprule
\textbf{} & \textbf{0 - 4 hours} & \textbf{4 - 12 hours} & \textbf{12 - 24 hours} & \textbf{24 - 48 hours} & \textbf{\textgreater 48 hours} \\ \midrule
\textbf{\# (\%)} & 32 (12.5\%) & 73 (28.6\%) & 78 (30.1\%) & 48 (18.8\%) & 24 (9.4\%) \\ \bottomrule
\end{tabular}
\caption{The participants' self-reported time spent on TikTok per week}
\label{tab:self_report}
\end{table}

\subsection{Data Collection}
We utilize a dataset~\cite{blind} containing a collection of 347 TikTok users' usage data and demographics. These users were asked to provide their downloadable data packages available from the TikTok mobile app, and to optionally participate in an additional survey to share their demographics and an estimate of self-reported time spent on TikTok. Users' video watch history was mandatory to donate in order to participate in the study; donation of additional data such as likes was incentivized but also optional.

\subsection{Data Selection}
Of the 347 TikTok users in the dataset, we excluded 92 (26.5\%) responses due to incompleteness or incompatibility with our analysis plan: 44 participants (12.7\%) who provided incomplete demographics, 34 participants (9.8\%) with a video watch history of one week or less, and 4 participants (1.2\%) of non-binary genders. While we recognize the importance of including non-binary individuals in our study, their representation was insufficient for generating statistically significant results. Our final dataset, therefore, consisted of data from 255 participants.

The demographics of participants in our dataset are gender-balanced: with 47\% of the participants identified as men. We classify age and education into two balance groups for each demographic topic (Age: 18-24, 25-64; Education: Below bachelor's degree, Holds a bachelor's degree or above), presented in Table \ref{tab:demographics}.

\subsection{Self-Reported Time}
Participants elected to report their self-reported usage time in an additional survey based on a six-point Likert scale. The two lowest value groups ("Less than two hours" and "Two to four hours") were merged to ensure balanced group sizes, presented in Table \ref{tab:self_report}. In designing the scale, we progressively increased the range of each time interval, i.e., instead of using an identical range throughout the scale. However, it allows heavier TikTok users to estimate their usage more accurately (i.e., to correctly select the appropriate range of use, as it is larger) and easily. We acknowledge this limitation and introduce an approach for achieving a more balanced representation in section~\ref{subsec:gap}.

\subsection{Measured Variables}
\label{ss:variables}
As noted in section~\ref{sec:related}, accuracy in self-report time estimations varies by different demographic, usage behaviors, and context-based factors. 
In our study, we examined these factors:

\subsubsection*{Demographics.}

We used these demographics to analyze the dataset: \textbf{age}, \textbf{education}, \textbf{gender}, and \textbf{``internet skills''}
\cite{Hargittai2005}. 
Participants' location data were excluded from our analysis due to an uneven distribution across groups: 137 (53.7\%) participants were from Africa, 84 (32.9\%) from North/Central America, 18 (7.1\%) from South America, 9 (3.5\%) from Europe, and 7 (2.7\%) participants did not disclose this information.

\begin{table*}[t]
\centering
\begin{tabular}{c|rrr|c}
\toprule
\textbf{\begin{tabular}[c]{@{}c@{}}Self-reported \\ Time\end{tabular}} & \multicolumn{1}{c}{\textbf{\begin{tabular}[c]{@{}c@{}}Under report\\  \# (\%)\end{tabular}}} & \multicolumn{1}{c}{\textbf{\begin{tabular}[c]{@{}c@{}}Over report\\  \# (\%)\end{tabular}}} & \multicolumn{1}{c}{\textbf{\begin{tabular}[c]{@{}c@{}}Accurate\\ \# (\%)\end{tabular}}} & \multicolumn{1}{c}{\textbf{\begin{tabular}[c]{@{}c@{}}Mean ABS \\ Error (hours)\end{tabular}}} \\ \midrule
0 - 4 hours           & 6 (18.8\%) & 0 (0\%)      & 26 (81.3\%)       & 1.14      \\
4 - 12 hours          & 4 (5.5\%)  & 53 (72.6\%)  & 16 (21.9\%)       & 2.72      \\
12 - 24 hours         & 0 (0\%)    & 75 (96.2\%)  & 3 (3.8\%)         & 9.45      \\
24 - 48 hours         & 0 (0\%)    & 48 (100\%)   & 0 (0\%)           & 21.56     \\
\textgreater 48 hours & 0 (0\%)    & 24 (100\%)   & 0 (0\%)           & 43.75     \\ \midrule
\textbf{Total}        & 10 (3.9\%) & 200 (78.4\%) & 45 (17.6\%)       & 12.00     \\ \bottomrule
\end{tabular}

    \caption{Accuracy metrics for the self-report times. We present the number of participants who under-report, over-report, and correctly report their actual logged usage. (Under-report/Over-report: the server-logged time is larger/smaller than the upper/lower bound of the self-report time category; Accurate: the server-logged time resides in the self-report time category. We also report the average absolute estimation error for each self-report time category by the number of hours.)
    }
    \label{tab:distribution}
\end{table*}

\subsubsection*{TikTok User Behavior.}

As this is the primary focus of our study, 
we evaluate the relationship between \textbf{logged time spent} and \textbf{self-reported time spent}. 
Following best practices from prior work~\cite{Ernala_2020}, we equally classify the server-logged time into four quartile groups. Durations falling under the lower quartile ($Q1$) were grouped together, as were those between the lower quartile ($Q1$) and the median ($Q2$), the median ($Q2$) and the upper quartile ($Q3$), and those exceeding the upper quartile ($Q3$).

We characterize other dimensions of activity on TikTok by measuring the number of \textbf{videos participants viewed per day} on average to collect another dimension of use. We also include \textbf{tenure} into the analysis, selected in light of observations from \citet{Burnell2021} that tenured and frequent users are likely to make more estimation errors. As TikTok does not expose the user's registration date in the donated data, we define tenure as the duration between the timestamp of the first video a participant watched, or their first recorded interaction with the TikTok ecosystem, and the time they donated their data. 

Additionally, it has been hypothesized that fragmented usage behaviors, which involve multiple activities scattered throughout the day, can lead to errors in self-reported time estimation \cite{schwarz2001asking}. Social media use is often fragmented \cite{voorveld2013age,voorveld2014investigating}, and prior works provide conflicting findings on whether fragmented social media use 
increases \cite{Ernala_2020} or decreases \cite{verbeij2021accuracy} time estimation errors. We measure this construct in our study by assessing the participants' number of \textbf{sessions} per day. We follow a previous work and define the beginning of a new session as when a participant starts watching a TikTok video after a minimum 300-second gap since their last use of the app \cite{Ernala_2020}. 

Finally, we measure the amount of \textbf{liked videos} per day participants had on average. Our choice to focus on ``likes'' was underpinned by wanting to explore forms of user engagement previously unstudied in time estimation research  
and motivated by psychological research indicating that perceived engagement, i.e., reporting on being emotionally engaged, affects time perceptions \cite{larson2006predicting}. Furthermore, most participants (99\%) shared their ``like'' data. 


\section{Results}

\begin{table}[t]
    \centering
    \begin{tabular}{lccc}
    \toprule

    \textbf{Coefficient} & \textbf{\begin{tabular}[c]{@{}c@{}}Self-Estimation \\ {Time}\end{tabular}} & \textbf{\begin{tabular}[c]{@{}c@{}}Over-Estimation \\ {Time (Gap)}\end{tabular}}   \\ \toprule
Age           & 0.23 & 0.23          \\
Gender        & 0.21 & 0.16          \\
Education     & 0.46**  {[}0.17, 0.75{]}& 0.51**  {[}0.20, 0.81{]}\\    
Like 2        & \begin{tabular}[c]{@{}c@{}}0.42*   {[}0.00, 0.85{]}\end{tabular}& 0.41      
                   \\
Like 3        & \begin{tabular}[c]{@{}c@{}}0.47*   {[}0.04, 0.91{]}\end{tabular}& \begin{tabular}[c]{@{}c@{}}0.47*   {[}0.01, 0.93{]}\end{tabular}      
                   \\
Like 4        & \begin{tabular}[c]{@{}c@{}}0.61*   {[}0.10, 1.11{]}\end{tabular}& \begin{tabular}[c]{@{}c@{}}0.60*   {[}0.07, 1.13{]}\end{tabular}       
                   \\

Internet Skills         & 0.23& 0.20         \\
Tenure        & -0.15& -0.06      \\
Videos          & 0.17 & 0.00      \\
Sessions      & \begin{tabular}[c]{@{}c@{}}-0.30*  {[}-0.56, -0.03{]}\end{tabular}& -0.28       
              \\
Logged Time Spent 2        & 0.00& 0.22             \\
Logged Time Spent 3         & -0.16 & 0.20           \\
Logged Time Spent 4         & 0.07 & -0.21           \\           
              \bottomrule          
    \end{tabular}
    \caption{Factors associated with estimates and confidence intervals in self-estimation and over-estimation time models.}
    \label{tab:estimation_model}
\end{table}

In this section, we detail the results of our analyses evaluating the accuracy of the time estimations of the 255 TikTok users in the dataset, and our efforts to identify factors that explain errors in their estimations. As a reference, we list their self-reports in Table \ref{tab:self_report}.

\subsection{Accuracy of Self-Reported Time} 
\label{subsec:correlation}

We first analyze how participants' self-report time estimates correlate with their observed usage history. As presented in Table \ref{tab:distribution},  a majority of the participants seemed to overestimate their time spent on TikTok. A minority of 45 users (17.6\%) report their usage correctly (their logged time matched the self-report time category). We observe that as the participants' self-reported time estimates increased, their accuracy decreased, and their average error increased.
 
To explore the factors that influence participants' time estimations, we run an ordinal logistic regression (specifically, a Cumulative Linked Model (CLM) using the ordinal R package~\cite{ordinalr}). This model enabled us to explore associations between the participants' demographics and usage characteristics and their self-reported time estimations.

In our model, we treated the self-reported time as an ordinal dependent variable, defined by the scale used in the survey. All other parameters (bolded in section \ref{ss:variables}) are independent variables. The number of likes and logged times are grouped by the quartiles and treated as categorical variables to avoid non-linear results. It is important to note, however, that the dataset had a 2-month gap in the ``like'' data, which was subsequently excluded from the calculation of average liked videos that might influence the validity of the results. We refer to this aspect in the Discussion section.  Education, gender, and age are also treated categorically. Internet skill is treated as a numerical variable. Tenure, number of videos, and number of sessions are similar numerical variables and subject to the natural logarithm ($ln$) transformation. We chose the lowest category of each ordinal or categorical factor as the baseline (e.g., youngest age and lowest quartile group of the number of likes), and for gender, we used men as the baseline category. The results of the models are summarized in the left column of Table~\ref{tab:estimation_model}.

We did not find a significant correlation between the server-logged time and participants' self-reported time estimates. However, we did find a significant negative correlation between the number of sessions and the self-reported time estimates. Additionally, participants with higher daily liked videos were more likely to overestimate their TikTok usage. Regarding demographics, higher education levels were linked with higher time spent estimates. Other factors did not show a significant correlation with self-reported time.

\subsection{Source of Overestimation}
\label{subsec:gap}
Given the inaccurate participants' self-estimations, we explored the factors related to their overestimations. We computed the gap between self-report time ($t_{rep}$) and server-logged time ($t_{log}$), and balanced the length of each scale group by multiplying the gap with the length of the scale that the server-logged time ($l_{rep}$) in, using the formula $t_{gap} = l_{rep} * |(t_{log} - t_{rep})|$. Participants with heavier usage estimations are ``punished'' with a higher gap as the scales of higher use estimations are larger and, therefore, are inertially easier to estimate accurately. For example, suppose the server-logged time of a participant is 10 hours while they reported that their usage was 12 - 24 hours. The correct category for the estimation is 4 - 12 hours, so their gap is $(12 - 4) * (12 - 10) = 24$.   We then segregated $t_{gap}$ by quartiles and used the same factors as in the estimation models with the same approach. The results are summarized in the right column of Table~\ref{tab:estimation_model}.

Similar to the estimation model, we see that education and the number of videos liked -- albeit in this model, only for those in the 3rd and 4th like quartile groups  -- positively correlate with a tendency to overestimate.
\section{Discussion and Limitations}
Our work investigates how users estimate the time they spend on the short-form video-centric social media platform TikTok, and explores what user demographics and engagement factors correlate with accurate time estimations. We find that participants generally overestimated their time spent on TikTok, and that server-logged time was not correlated with participants’ self-reported time. We see that participants who liked more videos were more likely to provide higher estimates of their time spent on TikTok, and had higher overestimation errors. Conversely, participants with more use sessions were more likely to provide lower time estimates and had lower overestimation errors.

Our results corroborate with prior work finding that users tend to overestimate their time spent using social media when asked to self-report \cite{coyne2023comparison,Ernala_2020,verbeij2021accuracy,Sewall2020,junco2013comparing,Boyle2022}. Prior work offers conflicting results on whether fragmented social media use -- using social media in several sessions per day -- results in more~\cite{Ernala_2020} or less~\cite{verbeij2021accuracy}
accurate estimations of time spent; our work corroborates with \cite{Ernala_2020} and finds that users with a higher number of sessions have lower error. Importantly, we also identify an engagement factor previously unstudied in time estimation research -- number of likes -- correlated with self-reporting inaccuracy.

This finding may relate to several prior results in psychology. While (i) engagement, measured by self-reported perceived engagement,  was related to reducing the perceived duration of activity \cite{larson2006predicting},  (ii) likes are discrete actions.  Psychological theory suggests that discrete events or actions may alter our perceptions of how long we have spent doing an activity~\cite{eagleman2008human} and (iii) that reflecting on our activities (e.g., in deciding whether to press the like button) may cause that activity to be perceived as taking a longer amount of time than reality~\cite{larson2006predicting}. 



Overall, we view our findings as additions to mounting evidence that self-reports of time spent on social media use are not accurate as stand-alone measures of actual social media use. To increase the rigor of future research on the implications of social media use, we must look beyond self-reports of time and into other forms of data collection, or other measurable variables. Actual logged data is an increasingly popular metric, but a collection of such data may involve technical errors
, high monetary costs for researchers, and ethical concerns.

With this in mind, assuming that future research will continue to lean on users' self-reports to some extent, we advocate for digital media researchers to identify other variables that accurately reflect participants' social media use. 
It is subject to curiosity whether
 the measurement of different dimensions of user engagement 
may supplement, or even supplant, self-reports of time spent on social media. 

As with every research, we note that our work has several limitations. First, our analysis is based on a relatively small dataset consisting of 255 TikTok users from various locations worldwide. Moreover, users' demographics (such as age and education) in the dataset are not comprehensively represented, compelling us to consolidate several categories to reach a balance for analysis. 
Nonetheless, our findings provide an initial exploration of the accuracy of self-estimation time spent on a short-form video-oriented social media platform.

Second, we acknowledge again that attaining logged data of users' social media usage is not a fool-proof measure \cite{parry2021systematic}. In our work, our dataset contained a technical error with the missing 2-months ``like'' data, potentially as a result of logging failures within TikTok infrastructure. This error influences 190 (74.5\%) participants. The missing data consists of approximately 20\% of the entire likes data, since the participants had an average number of 310 tenure days and a median of 304 days. While we assume that the users' behaviors are consistent and the missing period does not significantly influence participants' overall like data, it is worth noting that those participants whose usage history is largely concurrent with the missing period may be particularly affected and the calculated coefficient may diverge from the actual figure.

Future research interested in gauging social media use habits, instead of relying solely on self-reports of time, may ask participants questions about their engagement or use styles as additional proxies, based on the relationships we found. Variables like number of likes and sessions are also sometimes readily available to users within applications or through mobile phone tracking \cite{Ernala_2020, Ohme2020}, which adds reliability and cost-effectiveness to these types of self-reports. We hope our work will open a door to design a more accurate approach to estimating users' usage. 

\section{Conclusion}
We conduct the first study examining how TikTok users estimate their time spent on the app. We also assess the accuracy of their estimations using log data from user-donated data packages. Among the 255 participants in our dataset, we find the majority overestimated the time they spent using TikTok. Examining the participants by different dimensions of engagement, we see that participants with more active engagement (measured in ``likes'') provided higher estimations of their time spent on the app, and also had higher overestimation error. Conversely, we see that participants with a higher number of sessions of use had lower estimations of their time spent on the app, and lower overestimation error. 
Our findings build on prior work on the accuracy of people's social media usage self-reports, corroborating the notion that people generally overestimate their time spent on social media. We also highlight potentially relevant engagement factors to study in future work.
\newpage

\bibliographystyle{ACM-Reference-Format}
\bibliography{reference}
\appendix

\end{document}